\documentclass[a4paper]{jpconf}
\usepackage{graphicx}
\usepackage{siunitx}
\usepackage{amssymb}
\usepackage{bm}
\usepackage{color}

\begin{document}
\title{CePdAl - a  Kondo lattice with partial frustration}

\author{Veronika Fritsch$^1$, Stefan Lucas$^2$, Zita Huesges$^{2}$, Akito Sakai$^1$, Wolfram Kittler$^3$,
Christian Taubenheim$^3$, Sarah Woitschach$^2$, Bj{\o}rn~Pedersen$^4$, Kai~Grube$^5$, Burkhard Schmidt$^2$,
Philipp~Gegenwart$^1$, Oliver~Stockert$^2$, and Hilbert v. L\"{o}hneysen$^{3,5}$}

\address{$^1$Experimental Physics VI, Center for Electronic Correlations and Magnetism, Institute of Physics, University of Augsburg, 86135 Augsburg, Germany \\
$^2$Max Planck Institute for Chemical Physics of Solids, 01187 Dresden, Germany \\
$^3$Karlsruhe Institute of Technology, Physikalisches Institut, 76131 Karlsruhe, Germany \\
$^4$Heinz Maier-Leibnitz Zentrum (MLZ), Technische Universit\"{a}t M\"{u}nchen, 85748 Garching, Germany \\
$^5$Karlsruhe Institute of Technology, Institut f\"{u}r Festk\"{o}rperphysik, 76131 Karlsruhe, Germany}

\ead{veronika.fritsch@physik.uni-augsburg.de}

\begin{abstract}
Magnetic frustration, which is well-defined in insulating systems with localized magnetic moments, yields exotic ground states like spin
ices, spin glasses, or spin liquids. In metals magnetic frustration is less well defined because of the incipient delocalization of magnetic moments by the interaction with conduction electrons, viz., the Kondo effect. Hence,
the Kondo effect and magnetic frustration
are antithetic phenomena. Here we present experimental data of electrical resistivity, magnetization, specific heat and neutron diffraction on CePdAl, which is one of the rare examples of a geometrically frustrated Kondo lattice, demonstrating
that the combination of Kondo effect and magnetic frustration leads to an unusual ground state.
\end{abstract}

\section{Introduction}
Geometric frustration can prevent magnetic ordering despite the presence of a sizable magnetic exchange interaction. In insulating systems this leads to exotic ground states like spin ices \cite{Bramwell2001Spin}, spin glasses \cite{Mydosh1993Spin} or spin liquids \cite{Balents2010Spin}. The existing theoretical concepts to describe these states involve localized magnetic moments and usually their nearest and next-nearest neighbor exchange interactions without considering the charge degrees of freedom \cite{Ramirez1994Strongly}. In metals the situation is different due to the presence of conduction electrons coupling to the local moments.
Firstly, especially in $f$-electron systems, the indirect long-range Ruderman-Kittel-Kasuya-Yosida (RKKY) interaction renders the description by nearest and next-nearest neighbor exchange not in general sufficient. In $4f$-electron systems partial frustration is found in a few materials where one part of the magnetic moments forms long-range order, the other part remains disordered down to lowest temperature. Examples are systems like ${R}$InCu$_4$ with $R = $ Gd, Tb, Dy, Ho, Er and Tm \cite{Nakamura1993Anomalous,Fritsch2004Spin,Fritsch2006Correlation} or CePdAl to be presented here \cite{Doenni1996Geometrically}.
Secondly, due to the Kondo effect local moments can be screened by the conduction electrons, leading to a non-magnetic ground state. Magnetic frustration and Kondo effect can in this respect be regarded as being antithetic: The Kondo effect, yielding a delocalization of the magnetic moments
due to virtual transitions  of $4f$ electrons to the Fermi level, is not beneficial for the formation  of a frustrated state.  On the other hand, magnetic exchange interactions between the local moments can result in a breakdown of Kondo screening \cite{Senthil2004Weak}. Hence, the frustration parameter, the ratio between Curie-Weiss and N\'{e}el temperature as defined for insulating systems, is no longer useful in the presence of the Kondo effect.

CePdAl is a partially frustrated Kondo lattice with a Kondo temperature $T^{\ast} \approx 6\,\si{K}$ \cite{Doenni1996Geometrically,Goto2002Field,Woitschach2013Characteristic}. Therefore it provides the possibility to investigate the interplay between magnetic frustration and Kondo physics. Here we present measurements of the magnetization,
the electrical resistivity and the specific heat as well as experimental data from neutron scattering experiments indicating that the frustrated moments in CePdAl remain  fluctuating down to lowest temperature \cite{Oyamada2008Ordering} and affect the long-range order and magnetic excitations of the non-frustrated moments.

\begin{figure}
\begin{center}
\includegraphics[width=14pc]{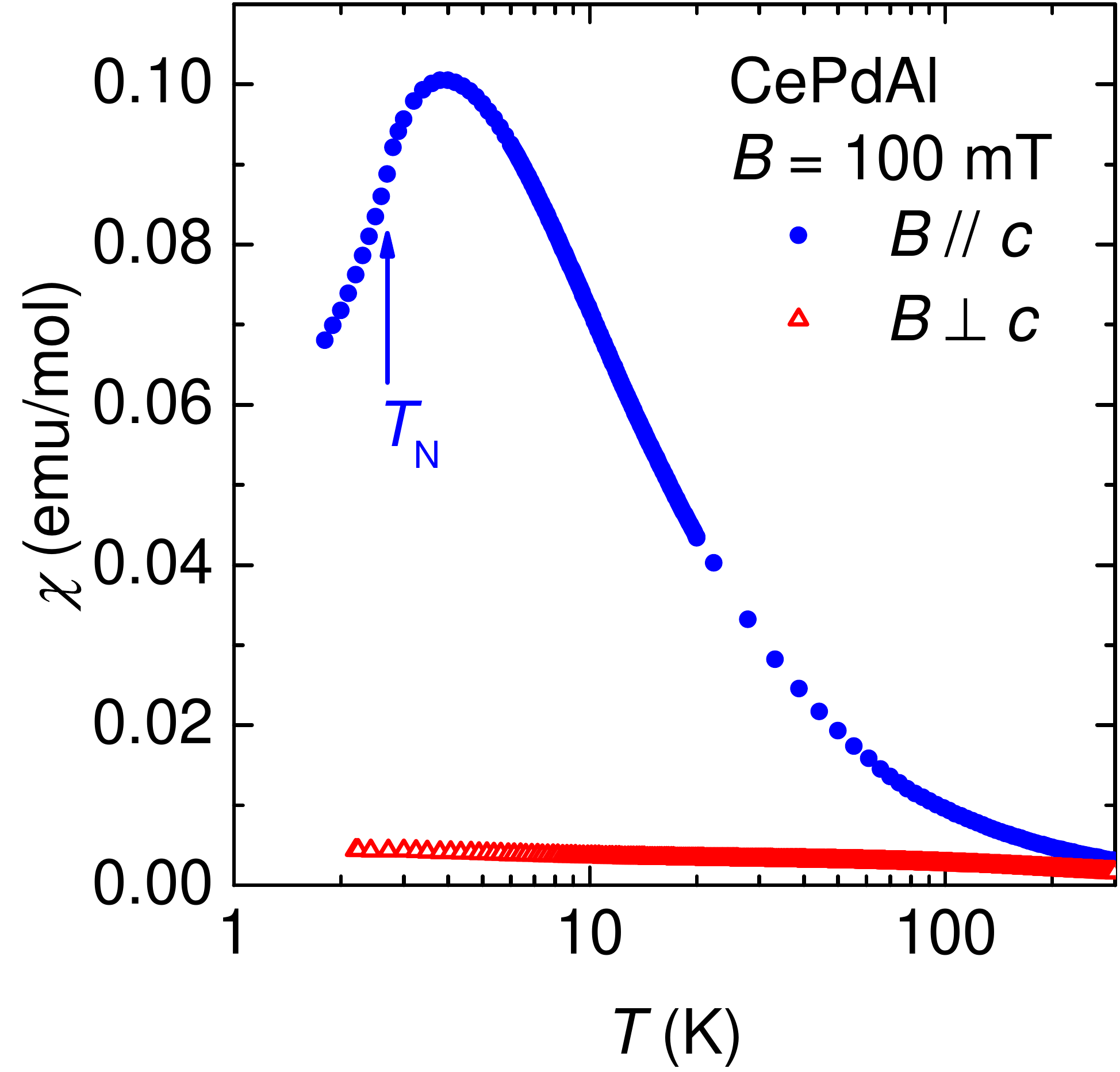}\hspace{4pc}
\begin{minipage}[b]{14pc}
\caption{DC susceptibility $\chi = M/B$ versus temperature $T$ in an external magnetic field $B = 100$~mT aligned parallel (blue circles) and perpendicular (red triangles) to the crystallographic $c$ axis.  The maximum of $\chi(T)$ indicates  antiferromagnetic order, while the maximum of $d\chi/dT$ yields the Ne\'{e}l temperature $T_N$ \cite{Fisher1962Relation}. \label{fig:chi}}
\end{minipage}
\end{center}
\end{figure}

\begin{figure}
\begin{center}
\includegraphics[width=14pc]{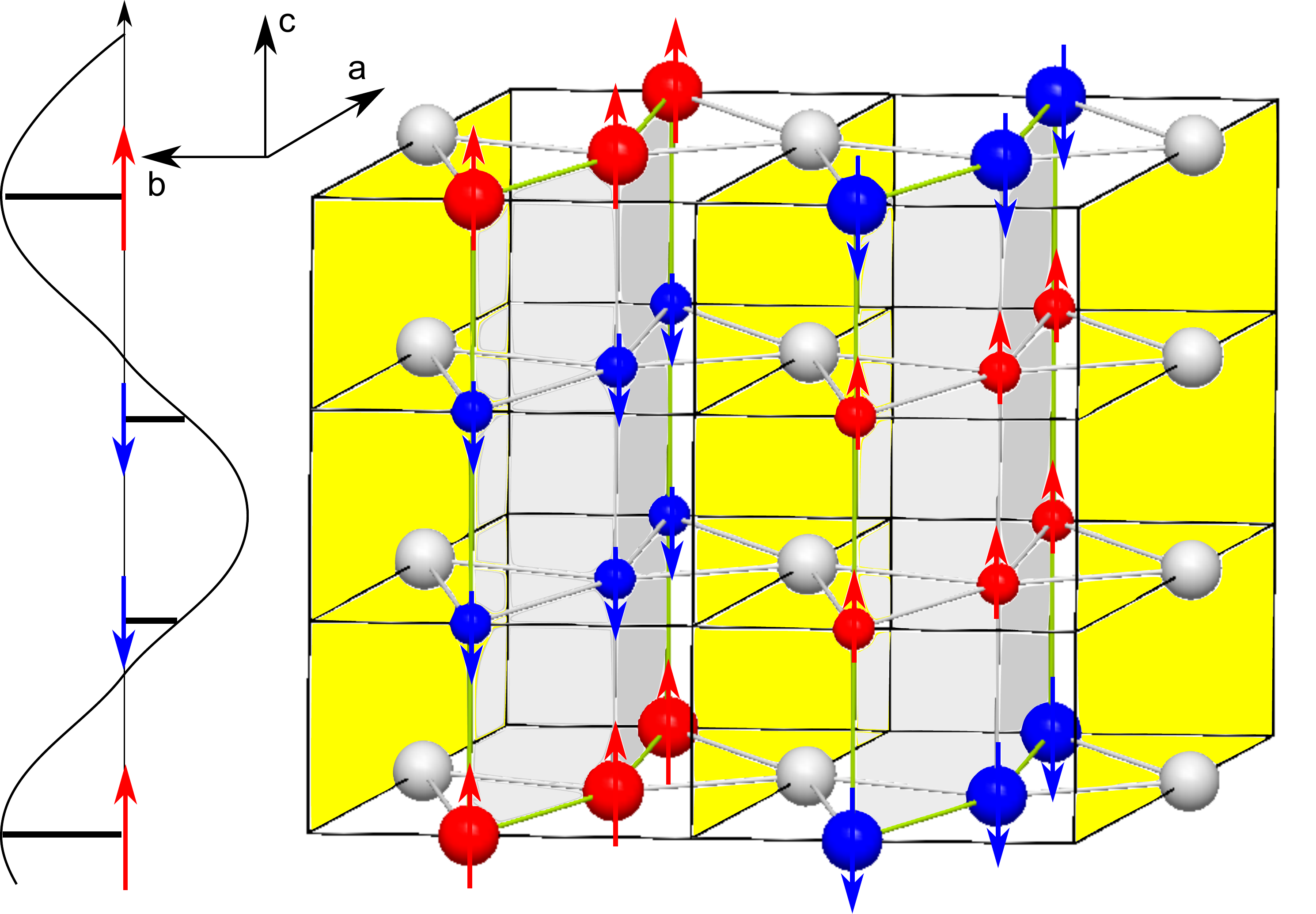}\hspace{4pc}
\begin{minipage}[b]{14pc}
 \caption{\label{fig:structure} Magnetic structure of CePdAl: two-thirds of the Ce moments form corrugated antiferromagnetic planes in the $ac$-plane with a sine-like modulation along the $c$ axis, which are separated by planes of frustrated moments (shown in yellow). The sketch neglects the small incommensuration $\tau^{\ast}$ along the $c$ direction}
\end{minipage}
\end{center}
\end{figure}

CePdAl crystallizes in the hexagonal ZrNiAl structure. The magnetic moments are located at the cerium sites, which form distorted kagom\'{e} planes stacked exactly on top of each other. At $T_{\rm N} = 2.7\,\si{K}$ antiferromagnetic order sets in. It is an Ising system with the magnetic moments aligned along the $c$ axis, as shown by the susceptibility measurements presented in Fig.~\ref{fig:chi}. These data are in nice agreement with previously published susceptibility
data by Isikawa et al. \cite{Isikawa1996Magnetocrystalline}, who attributed the magnetocrystalline anisotropy to the effect of the crystal electric field.
The magnetic structure derived from powder neutron diffraction with a wave vector $\bm{q} = \left(\frac{1}{2},0,\frac{1}{3}+\tau^{\ast}\right)$  \cite{Doenni1996Geometrically} is visualized in Fig.~\ref{fig:structure}. Within the basal plane, two thirds of the cerium moments form ferromagnetic chains, which are coupled antiferromagnetically and separated by the frustrated moments. Thus in contrast to the crystallographic structure with a single cerium site, the magnetic structure below $T_{\rm N}$ of CePdAl features three inequivalent cerium sites.

Due to the geometrical frustration present in the $ab$ plane in this compound one third of the magnetic moments does not participate in the long-range magnetic order \cite{Doenni1996Geometrically}. These moments form a rectangular lattice in the $ac$ plane (shown in yellow in Fig.~\ref{fig:structure}), separated from each other by antiferromagnetically ordered corrugated planes (grey shaded in Fig.~\ref{fig:structure}) \cite{Fritsch2014Approaching}.

\section{Experimental Details}
A single crystal  of CePdAl was grown by the Czochralski method \cite{Abell1989Handbook,Ames}.  A Physical Properties Measurement System (PPMS, Quantum Design) was used to obtain the specific heat and resistivity data in the temperature range between $350\,\si{mK}$ and room temperature. Specific-heat and resistivity measurements at lower temperatures  were performed  in a dilution refrigerator. The electrical resistivity was measured with a standard four-contact method employing a LR700 resistance bridge, for the specific heat measurements  a standard heat-pulse technique was used.
The magnetization measurements at $T = 0.5$ and $5$~K were performed in a Magnetic Properties Measurement System (MPMS, Quantum Design) equipped with a SQUID Magnetometer in magnetic fields up to $B = 7\,\si{T}$. The magnetization data at $T = 1.6$~K and the dc susceptibility between $1.6\,\si{K}$ and room temperature were measured in a Vibrating Sample Magnetometer (VSM, Oxford Instruments) in magnetic fields up to $B = 12$~T. Neutron scattering was performed at the instrument  RESI (FRM2, Munich) with a neutron wavelength  $\lambda =1.03\,\si{{\AA}}$ in a $^3$He cryostat.

\section{Results and Discussion}
\begin{figure}
\begin{center}
\includegraphics[width=14pc]{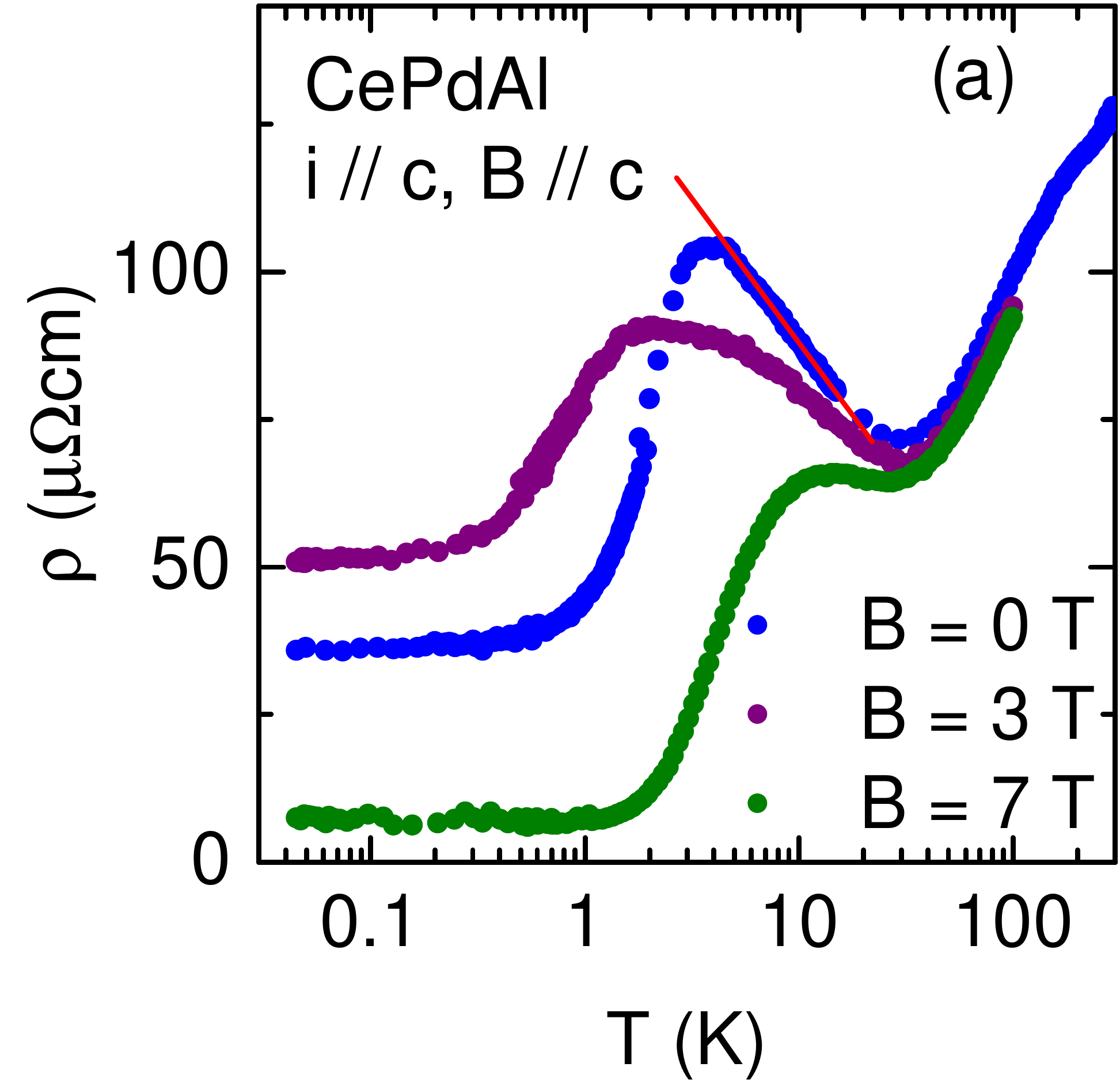}\hspace{4pc}\includegraphics[width=14pc]{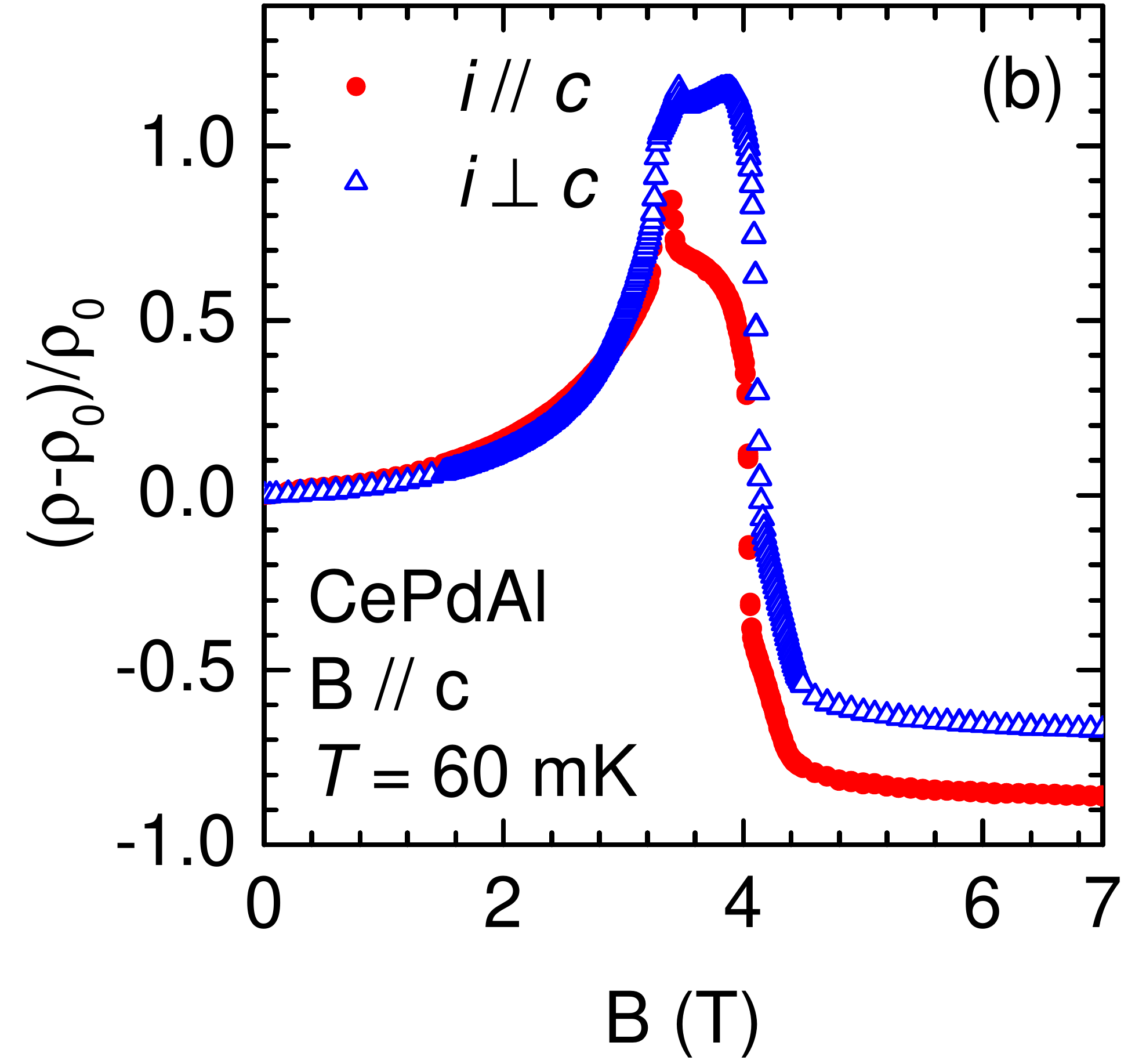}
\caption{\label{fig:widerstand} (a) Resistivity $\rho$ vs. temperature $T$ of CePdAl with the current $i$ applied along the $c$ axis in zero field (blue symbols) and an external magnetic field aligned parallel to the $c$ axis with $B = 3\,\si{T}$ (purple symbols) and $B = 7\,\si{T}$ (green symbols). The red line is a guide to eye, showing the logarithmic increase of the resistivity related to the Kondo effect. (b) Field dependence of the resistivity at $T = 60$~mK with the magnetic field aligned parallel to the $c$ axis and the current parallel (red circles) and perpendicular (blue triangles) to the applied field.}
\end{center}
\end{figure}

The electrical resistivity of CePdAl is shown in Fig.~\ref{fig:widerstand}~(a) in the temperature range between $40\,\si{mK}$ and $300\,\si{K}$ in zero field (blue symbols) and an external magnetic field applied parallel to the $c$ axis with $B = 3\,\si{T}$ (purple symbols) and
$B = 7\,\si{T}$ (green symbols). In zero field  a logarithmic increase of $\rho(T)$ (illustrated by the red line) is observed when lowering the temperature, characteristic for the presence of the Kondo effect. At $T_{max} \approx 4\,\si{K}$ the resistivity passes through a maximum and then drops steeply due to the onset of lattice coherence and antiferromagnetic order. A minimum around $T \approx 30\,\si{K}$ in the thermopower \cite{Huo2002Thermoelectric} confirms the presence of the Kondo effect in this material. An external magnetic field of $B = 3\,\si{T}$ lowers and broadens the maximum in $\rho(T)$ due to the lower N\'{e}el temperature $T_{\rm N} = 1.7\,\si{K}$, while the low-$T$ resistivity increases by roughly $30\%$ in $B = 3$~T.

In an external magnetic field $B = 7\,\si{T}$ the resistivity at the lowest attainable temperature  drops to $7.5\,\mu\Omega\,\si{cm}$ with the room temperature resistivity of $126\,\mu\Omega\,\si{cm}$ yielding a residual-resistivity ratio $RRR = 17$, which indicates a satisfying sample quality. Of course one could argue that magnetic impurities would be suppressed in an external magnetic field as well, however that would usually happen at significantly lower fields. The non-monotonic low-$T$ $\rho(T,B)$ is confirmed when considering the field dependence of the isothermal resistivity at lowest temperature, as presented in Fig.~\ref{fig:widerstand}~(b). The resistivity increases with increasing field with peaks at $3.4$~T and $3.9$~T (the latter being a shoulder for $i//B$) indicating magnetic transitions in nice agreement with the $\rho(B)$ data by Goto {\it et al.} \cite{Goto2002Field}, who also suggested that these transitions indicate the successive alignment of the  in zero field frustrated Ce moments with increasing magnetic field.

\begin{figure}
\begin{center}
\includegraphics[width=\linewidth]{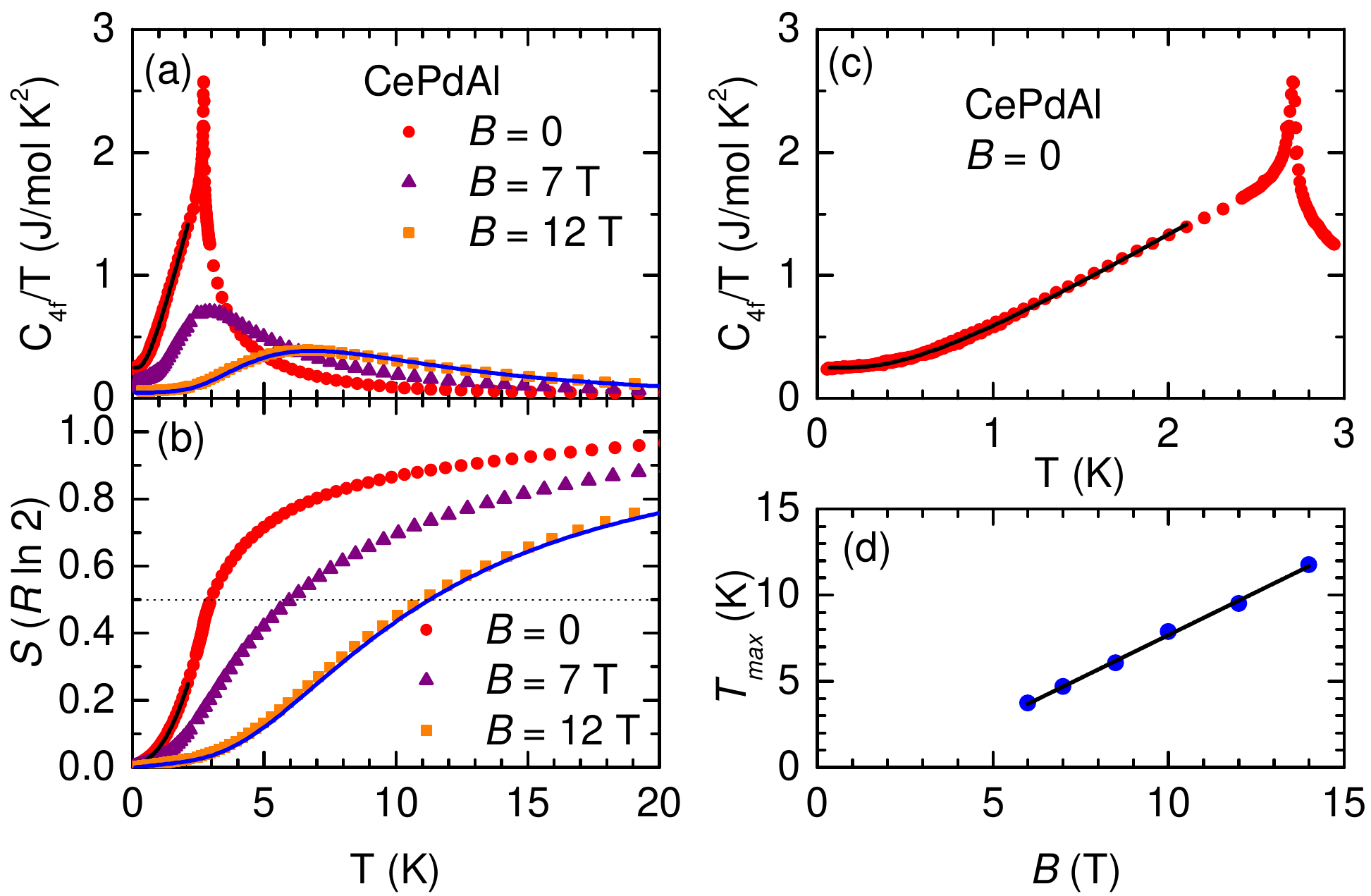}
\end{center}
\caption{\label{fig:spezHeat} (a) Specific heat of CePdAl in the representation $C/T$ vs $T$ in zero field (red circles) and in an external magnetic field along the easy axis $B = 7\,\si{T}$ (purple triangles) and  $B = 12\,\si{T}$ (orange squares). The black line is a fit to the zero-field data as described in the text. The  blue line is a fit with the Kondo model after ref.~\cite{Schotte1975Interpretation}, see text for details. (b)~Corresponding magnetic entropies $S(T)$. The blue line is a fit with the Kondo model as used in panel (a). (c)~Enlarged presentation of the zero-field specific heat and the corresponding fit.  (d)~Temperatures $T_{max}$ of the maxima of the Schottky anomaly in CePdAl vs magnetic field $B$.}
\end{figure}
In order to obtain insight into the magnetic low-energy states, we analyzed the specific heat of CePdAl in zero field down to very low temperatures.
Figure~\ref{fig:spezHeat}~(a) shows the $4f$-electron contribution $C_{4f}$ to the specific heat of of CePdAl in zero field, $B = 7\,\si{T}$ and $B = 12\,\si{T}$ with $B // c$.
$C_{4f}$ was obtained by subtracting the specific heat of the non-magnetic parent compound LuPdAl as well as the nuclear contribution $\propto 1/T^2$ due to the electric quadrupolar moments of Pd and Al calculated from high-field data \cite{Sakai2016Signature}. The black line, better visible in panel~(c) of Fig.~\ref{fig:spezHeat} is a fit of the zero field data with
\begin{equation}
 C_{4f}  = \gamma T + b\cdot T^2 \exp\left(-\frac{\Delta}{k_B T}\right)  \label{eq:Cfit}. \end{equation}
Here $\gamma T$ represents the electronic linear contribution to the specific heat with $\gamma = 250\,\si{mJ / mol K^2}$, which we attribute to the frustrated Ce moments. The second term $b\cdot T^2 \exp\left(-\frac{\Delta}{k_B T}\right)$ suggests the presence of gapped spin waves with $\Delta/k_B = 920\,\si{mK}$ in addition to the low-energy excitations due to frustrated Ce moments also found in previous NMR measurements \cite{Oyamada2008Ordering}. The  observed $T^2$ behavior is reminiscent of two-dimensional antiferromagnetic spin waves \cite{Ramirez1990Strong}, in agreement with the corrugated antiferromagnetic planes visualized in Fig.~\ref{fig:structure}.

\begin{figure}
\begin{center}
\includegraphics[width=14pc]{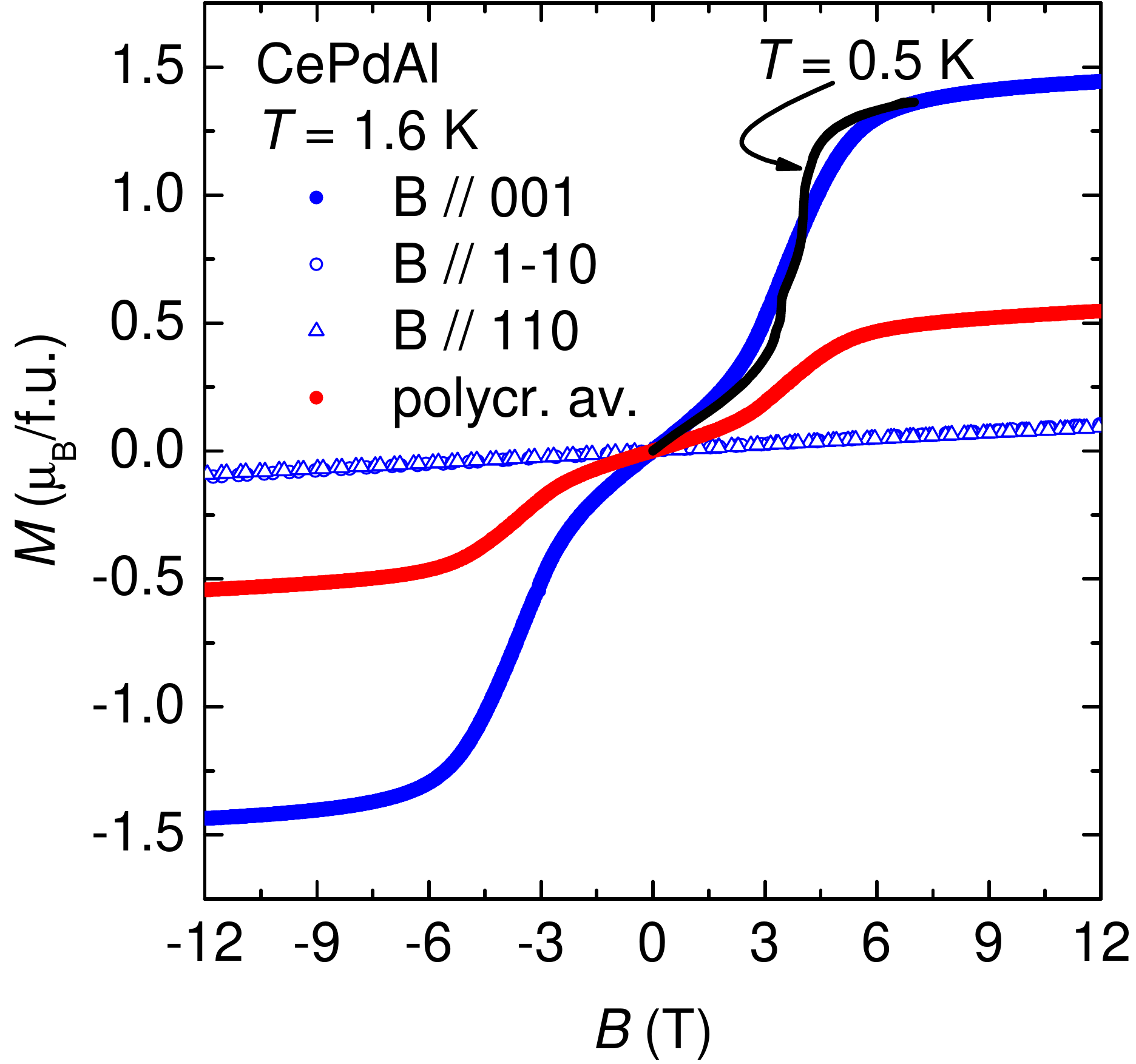}\hspace{4pc}
\begin{minipage}[b]{14pc}
\caption{\label{fig:magnetization} Magnetization of CePdAl at $T = 1.6$~K with the magnetic field aligned parallel to the $\left[001\right]$, $\left[110\right]$ and $\left[1\overline{1}0\right]$ direction. The red curve is the polycrystalline average of all three directions.  The black line shows the magnetization $M$ of CePdAl at $T = 0.5$~K with the magnetic field aligned parallel to the $\left[001\right]$.}
\end{minipage}
\end{center}
\end{figure}

The energy levels of the Ce are split into three doublets due to the local $m2m$ symmetry of the Ce ion. The crystal field levels were found at $\Delta/k_{\mathrm B} = 244$ and $510\,\si{K}$ by inelastic neutron scattering \cite{Woitschach2013Characteristic}. The ground state wave function is an almost pure $\left|\left.\frac{5}{2}\right>\right.$ doublet state, in line with the Ising anisotropy shown in the magnetization and susceptibility (see figs.~\ref{fig:chi} and \ref{fig:magnetization}). Thus at low temperatures we can assume an effective spin-$\frac{1}{2}$ system with an enhanced $g$-factor $g= 4.2$ to account for $J = \frac{5}{2}$ \cite{Abragam2012Electron33}.
The magnetic entropy calculated from the specific heat is shown in Fig.~\ref{fig:spezHeat}~(b).
For the zero-field measurement at the N\'{e}el temperature $T_{\rm N} = 2.7\,\si{K}$ less than $50\%$ of the expected entropy of $R \ln 2$ is gained, which is another hint to the frustration present in CePdAl.

As demonstrated by the specific heat data in $B = 7$ and $12$~T a Schottky anomaly evolves in magnetic fields. Fig.~\ref{fig:spezHeat}~(d) shows the temperature $T_{max}$ of the maxima in $C_{4f}$ of this Schottky anomaly in magnetic fields larger than $6$~T. The linear field dependence confirms that it is indeed a Schottky anomaly.
From the slope of the data we estimate the Zeeman splitting of the levels and the involved magnetic moment, the latter being $1.77\,\mu_{\mathrm{B}}$, which is in fair agreement with the ordered moment of $1.6\,\mu_{\mathrm{B}}$ found by D\"{o}nni {\it et al.} \cite{Doenni1996Geometrically} and compares well to $1.83\,\mu_\mathrm{B}$ found by Prok\v{e}s {\it et al.} \cite{Prokes2006Magnetic}.

The specific heat in $B = 12\,\si{T}$  can be described very nicely with the single-ion resonance-level Kondo model \cite{Schotte1975Interpretation}.
The solid blue lines in Fig.~\ref{fig:spezHeat}~(a) and (b) are fits with the resulting fit parameters as follows: Kondo temperature $T^{\ast} =  3.23 \pm 0.03\,\si{K}$ and Zeeman splitting $\Delta E/k_\mathrm{B} = 22.9\,\si{K}$, where the latter nearly perfectly agrees with $T_{max} = 9.5$~K when considering $k_B\cdot T_{max} = 0.42\cdot \Delta E$. In lower fields the emerging correlation effects compromise this fit by producing an additional background, which, however, does not shift $T_{max}$. Thus, although the magnetization data (see Fig.~\ref{fig:magnetization} and ref.~\cite{Goto2002Field}), suggest that in an external field $B = 7\,\si{T}$ aligned parallel to the $c$-axis the system is already in the paramagnetic state the specific heat in $B = 7$~T cannot be described with the single-ion resonance-level Kondo model. The presence of correlations between the magnetic moments, even in regimes beyond the long-range ordered state, is in line with the recent results of Prok\v{e}s {\it et. al} \cite{Prokes2015Probing}, who found that the three magnetic cerium sites still carry different magnetic moments at $T = 100\,\si{mK}$, $p = 0.85\,\si{GPa}$ and in $B = 9\,\si{T}$.

From the Sommerfeld coefficient $\gamma$ and the value of the magnetic susceptibility $\chi$ at low temperatures the Wilson ratio can be calculated \cite{Hewson1997Kondo}. The  Sommerfeld coefficient was determined from the value of $C_{4f}/T$ extrapolated to zero temperature to be $\gamma = 250$, $134$ and $57$~mJ/mol~K$^2$ in $B = 0$, $7$ and $12$~T. The magnetization, as  presented in Fig.~\ref{fig:magnetization}, was measured at $T = 1.6$~K in magnetic fields up to $B = 12$~T aligned along the easy axis $\left[001\right]$ and perpendicular along the hard axes $\left[110\right]$ and $\left[1\overline{1}0\right]$. The saturation moment at $B = 12\,\si{T}$ is $\mu = 1.44\,\mu_B/$f.u., roughly in line with the ordered moment $\mu \approx 1.6 \,\mu_B/$f.u. found in previous neutron diffraction experiments \cite{Doenni1996Geometrically} and our estimations from the specific heat data discussed above. In a high magnetic field with all moments aligned a saturation moment $m = 2.1\,\mu_B$ is expected for free Ce$^{3+}$. The significant reduction of our measured data is a further fingerprint of the Kondo screening effects in CePdAl.
For $B \perp c$ no transitions are found and $M/B$ is constant in field. From ref.~\cite{Isikawa1996Magnetocrystalline} we know that $M/B$ is temperature-independent below $T \approx 3$~K as well, so we can assume $\chi = dM/dB = 0.00712\,\mu_B\si{/mol\,T}$ for all temperatures below $3$~K and fields applied along the hard axis. For $B = 0$ we use our data taken at $T = 0.5$~K, which are in agreement with the data reported previously by Goto {\it et al.} \cite{Goto2002Field}. For the magnetic field aligned parallel to the easy axis the data at $T = 0.5$ and $1.6\,\si{K}$ above $B = 6$~K converge, thus we can use the data from our measurement at $T = 1.6$~K for $B = 7$~and $12$~T.  For the calculation of the Wilson ratio we take the polycrystalline average of these data.  The resulting values are $\chi = 0.039$, $0.019$ and $0.0078$~$\mu_B$/mol T in $B = 0$, $7$ and $12$~T.  The Wilson ratios calculated thereof are $R_W = 1.44$, $1.18$ and $1.17$ in $B = 0$, $7$ and $12$~T, which is close to the value for free electrons in all fields.
In a phenomenological ``two-component model'', which was for example used  previously for the analysis of CeCu$_{6-x}$Au$_x$ and CeAl$_2$, the specific heat at low temperatures is separated into two parts:  on the one hand magnetic excitations of the antiferromagnetic state, rapidly vanishing with decreasing temperature, and on the other hand heavy fermion excitations \cite{Schlager1993Magnetic,Bredl1978Specific}. In analogy one could argue that in zero field only the one third of frustrated, disordered  moments contributes to $\gamma$, which would result in a corrected Sommerfeld coefficient $\gamma' = 0.750$~J/mol~K$^2$ and a corrected Wilson ratio $R'_W = 4.32$. These values would qualify CePdAl as a strongly correlated heavy fermion system. In higher fields of $7$ and $12$~T this argument is not valid anymore, here the magnetic moments contribute equally to $\gamma$, resulting in a Wilson ratio as expected for normal metals.

\begin{figure}
\begin{center}
\includegraphics[width=14pc]{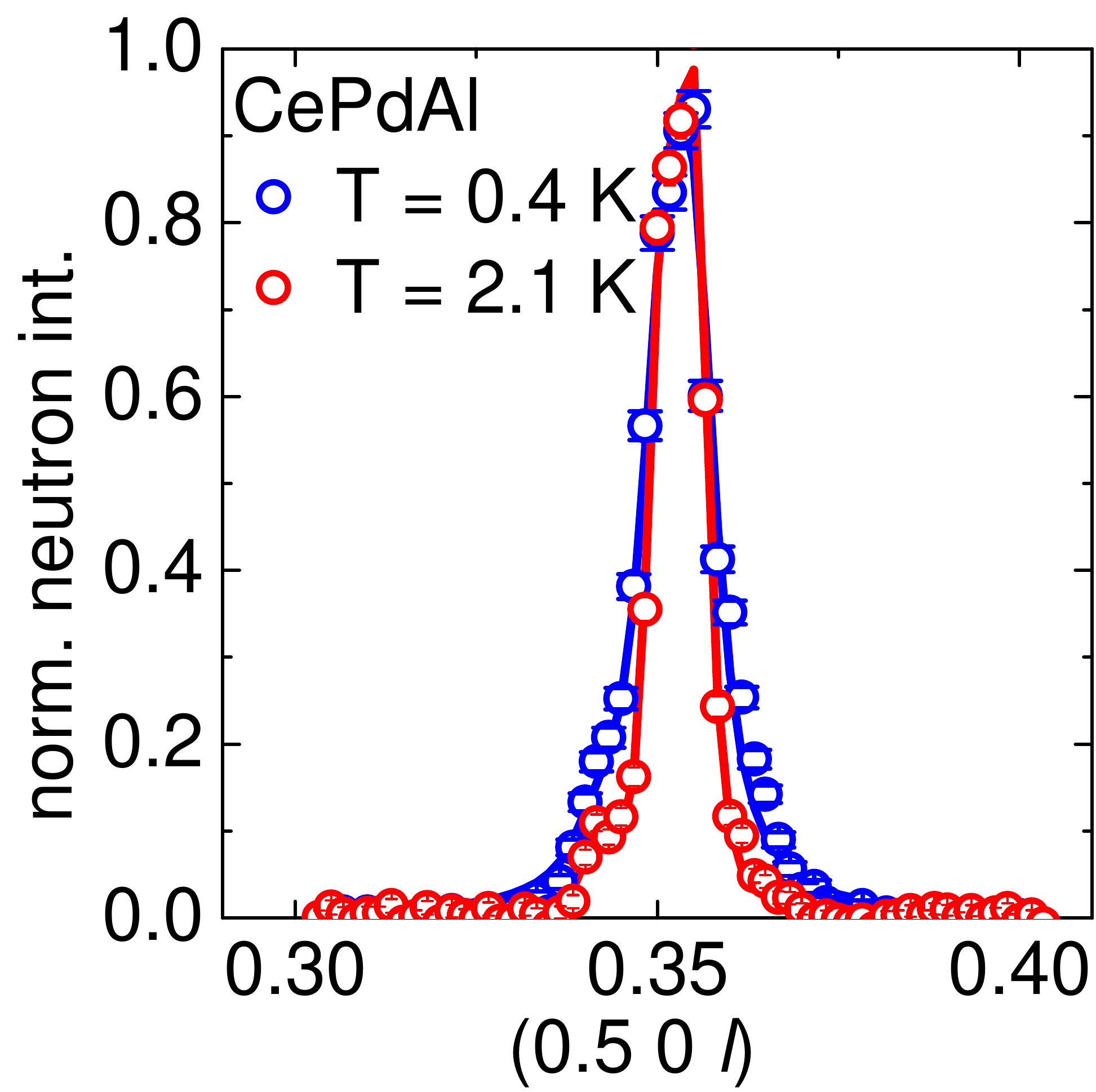}\hspace{4pc}
\begin{minipage}[b]{14pc}
\caption{\label{fig:neutrons}
Scans along $\left[001\right]$ across a magnetic Bragg $(0.5~0~0.35)$ peak at $T = 2.1$ and $0.4$~K.}
\end{minipage}
\end{center}\end{figure}

In order to take a closer look on the magnetic ground state of CePdAl we performed elastic neutron scattering experiments.
Figure~\ref{fig:neutrons} displays  scans across a magnetic Bragg position of CePdAl at $T = 2.1$ and $0.4$~K. The peak width of the magnetic Bragg peak increases towards lower temperatures, even well below $T_{\rm N}$. This measured width corresponds to antiferromagnetic domains with a size of the order of 200 {\AA} at $T = 0.4$\,K. Such a behavior is in marked contrast to usual magnetically ordered systems with true long-range order at lowest temperatures.
There must exist an effect limiting the size of the antiferromagnetic domains towards lower temperatures.   A possible origin of this behavior are the magnetically frustrated moments in CePdAl. Although no other long-range order is detected, one can assume that  the frustrated moments -  bearing a non-zero moment - couple to the ordered moments perturbing the long-range ordered structure resulting in a reduction of the  domain size.

\section{Conclusion}
Our results show clearly that magnetic frustration and the Kondo effect are present in CePdAl simultaneously. Our data suggest a bipartite system: on the one hand the antiferromagnetically ordered part, on the other hand the frustrated part.
However, these subsystems are not completely independent, as evidenced by the broadening of the magnetic Bragg peaks towards lower temperatures. Furthermore the missing entropy in magnetic fields above the magnetic transitions points to the presence of magnetic correlations and possibly magnetic frustration in areas beyond the long-range ordered regime. Overall CePdAl offers an excellent opportunity to explore the interplay between magnetic frustration and Kondo physics.

\section*{Acknowledgements}
This work was supported by the Deutsche Forschungsgemeinschaft through FOR 960, the Helmholtz Association through VI 521 and JSPS Postdoctoral Fellow for Research Abroad.

\section*{References}
\providecommand{\newblock}{}

\end{document}